\documentclass[twocolumn,amssymb,floatfix,deluxe,prd,showkeys]{revtex4-1}
\usepackage{graphicx,
longtable
}
\usepackage{dcolumn}
\usepackage{bm}

\newcommand{\NotreDame}{University of Notre Dame, Notre Dame, IN 46556-5670, USA}

\begin{document}
\title{Compact Neutrino Source}
\author{John LoSecco}
\affiliation{\NotreDame}
\date{\today}
\begin{abstract}
  Some evidence for sterile neutrinos has been found in short baseline
  observations where the measured neutrino flux did not agree with expectations.
  Systematic uncertainties from the expected values has limited the
  sensitivity of this approach.
  Observation at multiple distances can remove the
  normalization uncertainty by isolating the distance dependence.
  This doesn't work for high $\Delta m^{2}$ sterile neutrinos since they
  are fully mixed at most observation distances and only shift the normalization
  of the flux.  A compact intense source of neutrinos based on a subcritical
  fission reactor would permit observation
  of oscillations on submeter distance scales and clearly distinguish between
  a systematic normalization and the $L/E$ dependence expected from
  oscillations.
\end{abstract}
\maketitle
Hints for the existence of sterile neutrinos mixing with electron and muon
neutrinos come from several observations \cite{Kaiser}.  Evidence comes both
from appearance, when the source is a muon neutrino and from observed event
rates being below expectations, when the source is an electron neutrino.
If one uses a neutrino oscillations framework to model the observations
the short distances involved suggest a large mixing mass and the low
transition rates suggest a small mixing angle.  In this model the transition
rate for electron neutrinos can be approximated by:
\[
P_{e \rightarrow e}=1-\sin^{2}(2 \theta_{es}) \sin^{2}(\Delta_{es})
\]
where $\Delta_{es} = 1.267 ( m^{2}_{e} - m^{2}_{s} ) \frac{L}{E} = 1.267
\Delta m^{2}_{es} \frac{L}{E}$.
This is a reasonable approximation when $\Delta_{es} >> \Delta_{ij}$.
Where $\Delta_{ij}$ is the scale of the known three flavor oscillations.
Suggested values for $\Delta m^{2}_{es}$ are about 1 $eV^{2}$ and
$\sin^{2}(2 \theta_{es})=0.1$ \cite{Kaiser}.  So the approximation is good
for short
baseline experiments.  The small value of $\sin^{2}(2 \theta_{es})$ suggests
that high statistics are needed to see the effect and that systematic
errors may dominate the measurement.  Values as high as
$\Delta m^{2}_{es} > 5$ eV$^{2}$ can be fit \cite{Kaiser}.

The interaction of MeV scale antineutrinos (such as from fission products) on
hydrogen, $\bar{\nu_{e}} + P \rightarrow e^{+} + N$
is a well understood process that
lends itself to the background resistant delayed coincidence method where the
neutron is measured after the initial neutrino interaction occurs to produce
a positron.  Almost all of the neutrino energy is carried by the positron
so neutrino energy ambiguity is determined by detector resolution and not
neutrino reaction kinematics.

The transition probability depends on the distance from the neutrino source
to the detection point.  Even if the detector has good spatial resolution
one must average over the source size.
\[
\frac{1}{L} \int \sin^{2}(1.267 ( m^{2}_{e} - m^{2}_{s} ) \frac{L}{E}) dL \approx \frac{1}{2}
\label{Av}
\]
If the source size is comparable to
about $\frac{1}{4}$ of the oscillation length
($L_{Osc}=\frac{2 \pi E}{1.267 \Delta m^{2}_{es}}$) or larger averaging over the
source size is equivalent to shifting the flux normalization.  This leads to an
ambiguity between the sterile neutrino hypothesis and the normalization
of the source flux.  It is advantageous to have a small source size.

A short baseline experiment from an intense compact low energy antineutrino
source is an ideal tool to study sterile neutrinos.  $^{144}Ce$ has been
chosen as
a compact neutrino source for such an experiment \cite{CeSOX}.
The end point energy of the $^{144}Pr$ daughter beta decay is about 3 MeV.
The difficulty of creating, transporting and maintaining such a source
has encouraged a search for alternatives.
\begin{figure}
\includegraphics[width=0.48\textwidth]{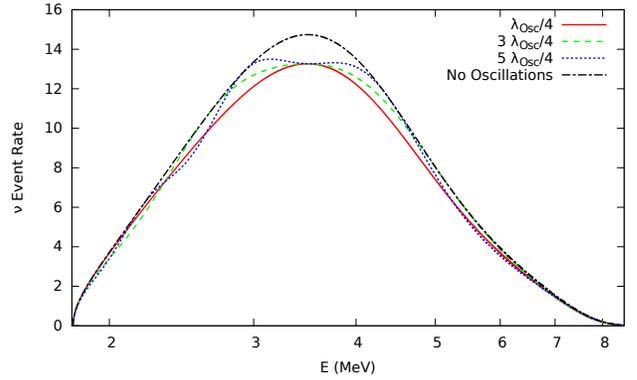}
\caption{\label{SterileOsc} The expected fission product neutrino event
  spectrum at various distances from a compact $^{239}Pu$ source.  The
  oscillation length is
  $L_{Osc}=\frac{2 \pi E}{1.267 \Delta m^{2}}$.  The figures are made with
  $L_{Osc}$ computed at $E=3.5$ MeV, near the peak rate and for
  $\sin^{2}(2 \theta_{es})=0.1$.}
\end{figure}

In reference \cite{Janet} it is suggested that cyclotrons be used to produce
beta unstable radionuclides such as $^{8}$Li near a large mass detector
so the transportation issues associated with radioactive sources is mitigated.
The end point energy of the $^{8}$Li beta decay is about 16 MeV.
The $^{8}$Li is made {\em in situ} from neutron capture on $^{7}$Li.
The interaction length for neutron capture determines the source size.
For an isotopically pure target of $^{7}$Li metal the neutron capture length
is about 475 cm.  The presence of $^{6}$Li is problematic since its neutron
capture cross section is 21 times larger but it yields no neutrinos.

Application of accelerator technology to produce subcritical fission power
sources \cite{Carlo} suggests a more efficient way to generate an intense
compact neutrino beam.  In the subcritical fission reactor, accelerator induced
fission greatly amplifies the power injected and produces the neutron rich
beta decay unstable fission products that are responsible for the antineutrinos
produced at nuclear power stations.  Unlike self sustaining chain reactions
where the geometry and design require a large core, a subcritical source has
some advantages.  For example, cooling the source could be engineered to permit
much shorter distance between the neutrino source and the detector.  This
minimum being determined by radiation shielding requirements.  Backgrounds
which are not beam associated can be measured by turning off the accelerator
hence removing the neutrino source.  Removing the source of neutrinos is
rarely practical for commercial nuclear reactor operations.

In reference \cite{Carlo} most of the fission power comes from $^{233}U$
bred {\em in situ} from $^{232}Th$ but reactors utilizing $^{239}Pu$ bred
from $^{238}U$ are also possible.  For a neutrino source it also might be
reasonable to dispense with the breeding phase and start with a subcritical
assembly of $^{233}U$, $^{235}U$ or $^{239}Pu$.  Due to the high neutron
fission cross section for these materials interaction lengths for pure metals
is below a millimeter.  The actual source size is determined by safety and
engineering issues.

A low power compact fission source might be able to dispense with the
high power accelerator and provide neutrons another way.  About $3\times10^{13}$
thermal neutrons per second are needed to sustain 1 kW of fission power,
enough to probe the short distances of the highest values
of $\Delta m^{2}_{es}$.

Figure \ref{SterileOsc} illustrates the spectral distortion of a compact
$^{239}Pu$ fission product neutrino source due to sterile
neutrinos for several different source to detector distances.  The peak
in the unoscillated event rate is at about 3.5 MeV.

Fission product neutrino sources are well studied and understood.  In fact
discrepancies \cite{Reactor} at modest distances from nuclear reactors
between the measured neutrino rate and the expected one is one hint for sterile
neutrinos.

A compact source would permit the L/E measurement and distinguish
normalization errors from a true neutrino oscillation.  An L/E measurement
could do this with no dependence on the calculated neutrino source error by
measuring simultaneously at different distances from the compact source.

The rate of neutrino interactions in a liquid scintillator detector ($CH_{2}$)
is about
76 events per ton-day 10 meters from a 10 MW source.  We have used the neutrino
spectrum estimates for $^{239}Pu$ from reference \cite{NuFlux} and the cross
section from reference \cite{VogBeac}.

The neutrino spectrum may change with time as the isotopic content of the
target and its spatial distribution changes.  But as long as spectra at
all baselines are compared to the same exposure, fuel evolution issues are
restricted to integrating over the source.  If the source is compact it is
less sensitive to inhomogeneities due to burn up.

Since the oscillation length is a function of the neutrino energy
($L_{Osc}=\frac{2 \pi E}{1.267 \Delta m^{2}_{es}}$) one can reduce the
effect of the neutrino source size by using a higher neutrino energy.
But the higher neutrino energy requires longer baselines which reduces
the neutrino flux.  Higher neutrino energy also means that additional detection
channels will open up making energy resolution an issue as more of the neutrino
energy gets transferred to nuclear excitation and neutron emission.
There are also fewer convenient sources for higher energy neutrinos.

There are cosmological observations \cite{Planck2015} which are in tension with
the possible existence of sterile neutrinos ($N_{eff} = 3.2 \pm 0.5$ and
$\Sigma m_{\nu} < 0.32$ eV).

I would like to thank Boris Kaiser, Thierry Lasserre, Bill Louis and
Carlo Rubbia for some helpful remarks.

\end{document}